\documentclass[final,1p,times]{elsarticle}
\usepackage{array}
\usepackage{multirow}
\usepackage{subcaption}
\usepackage{floatrow}
\newfloatcommand{capbtabbox}{table}[][\FBwidth]

\usepackage{amssymb}

\journal{Knowledge-Based Systems}

\begin{document}

\begin{frontmatter}



\title{Complex network modelling of EEG band coupling in dyslexia: an exploratory analysis of auditory processing and diagnosis.}


\author[inst1]{Nicolás J. Gallego-Molina}
\ead{njgm@ic.uma.es}
\cortext[cor1]{Corresponding author.}
\author[inst1,inst3]{Andr\'es Ortiz\corref{cor1}}
\ead{aortiz@ic.uma.es}
\author[inst2,inst3]{Francisco J. Mart\'inez-Murcia}
\author[inst1,inst3]{Marco A. Formoso}
\author[inst4]{Almudena Giménez}

\affiliation[inst1]{organization={Department of Communications Engineering, University of Malaga},
            country={Spain}}

\affiliation[inst2]{organization={Department of Signal Theory, Telematic and Communications, University of Granada},
            country={Spain}}
\affiliation[inst3]{organization={Andalusian Data Science and Computational Intelligence Institute (DasCI)},
            country={Spain}}
\affiliation[inst4]{organization={Department of Developmental Psychology, University of Malaga},
            country={Spain}}

\begin{abstract}
Complex network analysis has an increasing relevance in the study of neurological disorders, enhancing the knowledge of brain’s structural and functional organization. Network structure and efficiency reveal different brain states along with different ways of processing the information. This work is structured around the exploratory analysis of the brain processes involved
in low-level auditory processing. A complex network analysis was performed on the basis of brain coupling obtained from Electroencephalography (EEG) data, while different auditory stimuli were presented to the subjects. This coupling is inferred from the Phase-Amplitude coupling (PAC) from different EEG electrodes to explore differences between controls and dyslexic subjects. Coupling data allows the construction of a graph, and then, graph theory is used to study the characteristics of the complex networks throughout time for controls and dyslexics. This results in a set of metrics including clustering coefficient, path length and small-worldness. From this, different characteristics linked to the temporal evolution of networks and coupling are pointed out for dyslexics. Our study revealed patterns related to Dyslexia as losing the small-world topology. Finally, these graph-based features are used to classify between controls and dyslexic subjects by means of a Support Vector Machine (SVM).
\end{abstract}

\begin{keyword}
Dyslexia diagnosis \sep EEG \sep Complex network \sep Graph analysis \sep  PAC \sep SVM
\end{keyword}

\end{frontmatter}


\section{Introduction}
\label{sec:Introduction}
Based on graph theory, analysis of complex networks has been used in many fields related to social sciences, physics and information technology. The advances achieved have been translated to neuroscience and applied to the analysis of complex networks originated in the brain. Furthermore, it provides a powerful way of quantifying the brain's structural and functional systems \cite{bullmore_complex_2009}. These complex networks rely on data obtained with brain mapping techniques such as diffusion MRI, functional MRI (fMRI), EEG and MEG. This way, complex network analysis helps to quantify brain networks with a small number of neurobiologically meaningful and easily computable measures. This is a useful setting for exploring structural–functional connectivity relationships of the brain processes. Therefore, it is a promising mechanism to reveal connectivity abnormalities in neurological and psychiatric disorders \cite{rubinov_complex_2010}.

There are many examples in the literature of using complex network analysis to investigate the brain network characteristics. For example, the analysis of complex networks from physiological and pathological brain aging such as Alzheimer’s disease in \cite{stam_smallworld_2006,vecchio_human_2014,ortiz_exploratory_2015,munilla_construction_2017}. There are other studies that employ functional connectivity and complex network analysis for other neurological disorders as epilepsy and schizophrenia. In   \cite{leitgeb_brain_2020}, complex networks are used to detect changes in EEG-based functional connectivity patterns in six pediatric patients with childhood absence epilepsy. Other works as  \cite{rutter_graph_2013} apply graph theoretical analysis to MEG functional connectivity networks to study schizophrenia.

In this context, we present some existing studies about functional network connectivity and organization in Developmental Dyslexia (DD). The use of EEG signals is widely extended. In  \cite{fragagonzalez_graph_2016}, the authors examined differences in the topological properties of functional networks using graph analysis. Also, EEG connectivity analysis was used in  \cite{zaric_altered_2017} to investigate whether dysfunctional connectivity scales with the level of reading dysfluency. In this field, Phase-Amplitude Coupling was calculated in  \cite{power_neural_2016} for measuring the accuracy of low-frequency speech envelope encoding. Another powerful tool for characterizing brain networks and how they are disrupted in neural disorders (in this case Dyslexia) are functional connectivity analyses of fMRI data as in \cite{finn_disruption_2014,edwards_dyslexia_2018,qi_more_2016,bailey_applying_2018}. Despite methodological heterogeneities, the analysis of brain networks reveals a general trend towards highly clustered and highly integrated structures, consistent with a small-world network topology across studies\cite{sporns_small_2006}.

DD is a neurological condition that causes learning disability disorders and affects between 5$\%$ and 13$\%$ of the population \cite{peterson_developmental_2012}. DD diagnosis is a relevant area of study that benefit from applying complex network analysis. Within this topic, the early diagnosis is receiving increasingly attention. It is an essential task to help dyslexic children have a proper personal development with the application of preventive strategies for teaching language. For this purpose, EEG signals enable the use of specific measures not related to reading for an objective and early diagnosis in pre-readers subjects. In the present work, complex network analysis is used to examine the different network characteristics between dyslexics and controls. This analysis relies on Phase-Amplitude Coupling based features and it is focused on the exploratory analysis of the brain processes.  With that in mind, the relations described with PAC between frequency bands are used to construct the networks. Thus, including information of the PAC intra-electrode through time for every subject. To this end, PAC measured and graph theory provide a basis to analyze and quantify the differences. Finally, this network characteristics are used as features to train a Support Vector Machine classifier.

The paper is organized as follows. Section \ref{sec:Methods} presents details of the database, describes the auditory stimulus and the methods used. Then, Sections \ref{sec:Results} and \ref{sec:Discussion} state and discuss the experiments results, and finally, Section \ref{sec:Conclusions} draws the main conclusions and the future work.

\section{Materials and Methods}
\label{sec:Methods}

In this work, graph analysis is the principal tool to study the temporal evolution of the DD complex systems. This analysis is based on the EEG data provided by the Leeduca Study Group at the University of Málaga \cite{ortiz_dyslexia_2020}. Forty-eight participants took part in the study by the Leeduca Study Group, including 32 skilled readers (17 males) and 16 dyslexic readers (7 males) matched in age (t(1) = -1.4, p $>$ 0.05, age range: 88-100 months). The mean age of the control group was 94.1 $\pm$  3.3 months, and 95.6 $\pm$ 2.9 months for the dyslexic group. The experiment was carried out with the understanding and written consent of each child’s legal guardian and in the presence thereof. 

EEG signals were recorded using the Brainvision actiCHamp Plus with 32 active electrodes (actiCAP, Brain Products GmbH, Germany) at a sampling rate of 500 Hz during 15-minutes sessions, while presenting an auditory stimulus to the subject. Participants underwent 15 minute sessions in which they were presented white noise stimuli modulated at 4.8, 16, and 40 Hz sequentially in ascending and descending order, for 2.5 minutes each. All participants were right-handed Spanish native speakers with no hearing impairments and normal or corrected–to–normal vision. Dyslexic children in this study have all received a formal diagnosis of dyslexia in the school. None of the skilled readers reported reading or spelling difficulties or have received a previous formal diagnosis of dyslexia. The locations of 32 electrodes used in the experiments is in the 10–20 standardized system.

\subsection{Signal preprocessing} 
EEG signal preprocessing constitutes an important stage, due to the low signal-to-noise ratio of EEG signals and the number of artifacts presents in the signals. The preprocessing stage includes the following steps:  first, blind source separation with Independent Component Analysis (ICA) was used to remove artifacts corresponding to eye blinking signals in the EEG signals. Then, the signal of each channel was normalized independently to zero mean and unit variance and referenced to the signal of electrode Cz. Baseline correction was also applied. Finally, signals were segmented into 15.02s long windows to analyze PAC temporal patterns correctly \cite{dvorak_proper_2014}. This is the optimal window length for which a sufficiently high number of slow oscillation cycles are analyzed. Shorter windows lead to overestimates of coupling and lower significance. According to \cite{dvorak_proper_2014}  the window length is one of the main requisites for robust PAC estimation. Moreover, it is a key concept for a correct statistical validation of the results. On this purpose, surrogate tests were used without needing long data windows that assumes stationarity of the signals within the window.

\subsection{Phase-Amplitude Coupling}
\begin{figure}[!htb]
\includegraphics[width=\textwidth]{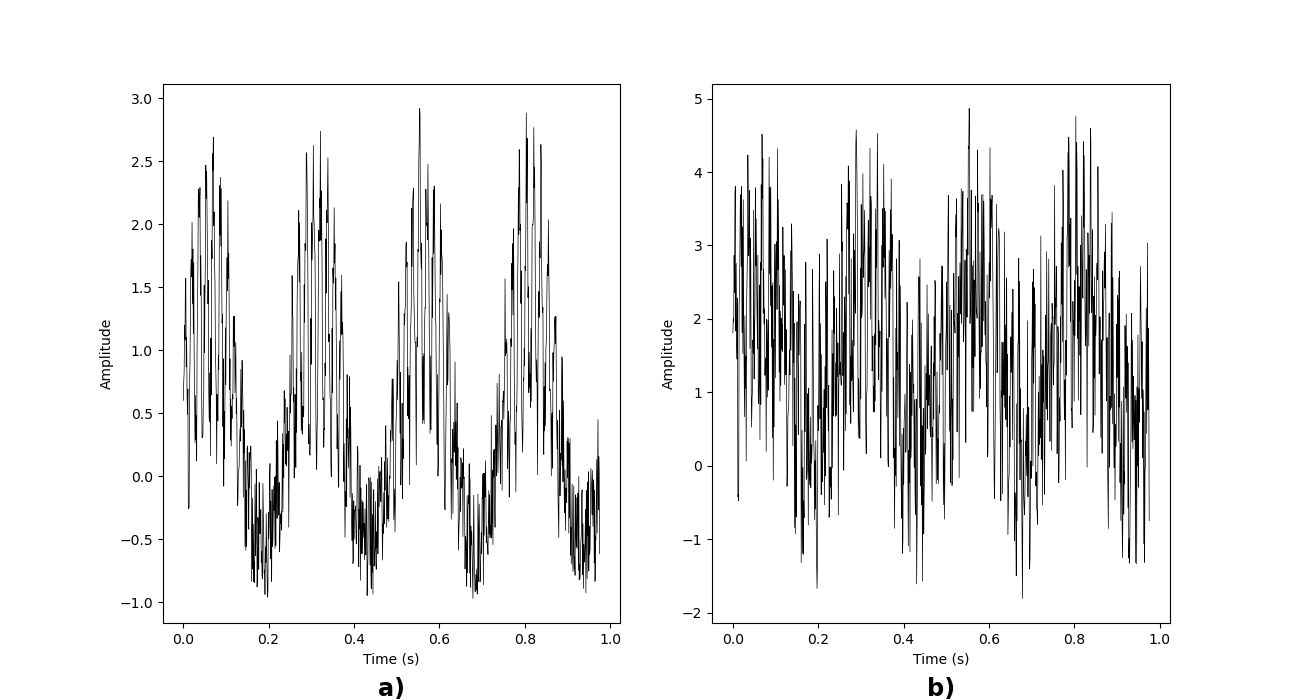}
\caption{Synthetic signals containing PAC between the phase of the band Theta (4 Hz) and the amplitude of the band Gamma (60 Hz). a)Strong coupling b)Weak and noisy coupling} \label{fig1_}
\end{figure}

The representation of a complex system as a network requires the definitions of the nodes and edges that form the graphs. In this work, the edges correspond to the PAC measured between a pair of frequency bands. This type of Cross Frequency Coupling (CFC) has been proposed to coordinate neural dynamics across spatial and temporal scales \cite{aru_untangling_2015}. According to Canolty et al. \cite{canolty_functional_2010} CFC have a functional role representing the interaction between different brain rhythms. It serves as a mechanism to transfer information from large scale brain networks (where low-frequency brain rhythms are dynamically entrained) to local cortical processing reflected by high-frequency brain activity. Furthermore, CFC has a potential relevance for understanding healthy and pathological brain function. In particular, Phase-Amplitude CFC has received significant attention \cite{dvorak_proper_2014,vandermeij_phaseamplitude_2012} and play an important functional role in local computation and long-range communication in large-scale brain networks \cite{canolty_functional_2010}. It is described as the modulation of the amplitude of the high frequency oscillation by the phase of the low frequency component. Fig. \ref{fig1_} represents an example of two synthetic signals that contain PAC between the phase of the  Theta band (4 Hz) and the amplitude of the Gamma band (60 Hz). The signal on the right show a weaker and nosier coupling.

As we mentioned above, we measured the PAC as the coupling between the phase of a slower oscillation and the amplitude of a faster oscillation. In particular, the amplitude modulation of the Gamma (30–100) Hz and Beta (12-30) frequency bands by the phase of the Delta (0.5-4) Hz, Theta (4-8) Hz, Alpha(8-12) Hz bands. Additionally, the phase of Beta band for the amplitude of Gamma band. There are different PAC descriptors as presented in \cite{hulsemann_quantification_2019}: Phase-Locking Value, Mean Vector Length (MVL), Modulation Index (MI), and Generalized-Linear-Modeling- Cross-Frequency-Coupling. In this way, there is no convention yet of how to calculate phase-amplitude coupling. Here, the modulation index (MI) is used, as defined  by Tort et al. \cite{tort_dynamic_2008,tort_measuring_2010}. It is considered in \cite{tort_measuring_2010} as one of the best performance methods for assessing the intensity of
the PAC and it is well suited for the characteristics and length of the data used in the present work.

\begin{figure}[!htb]
\includegraphics[width=\textwidth]{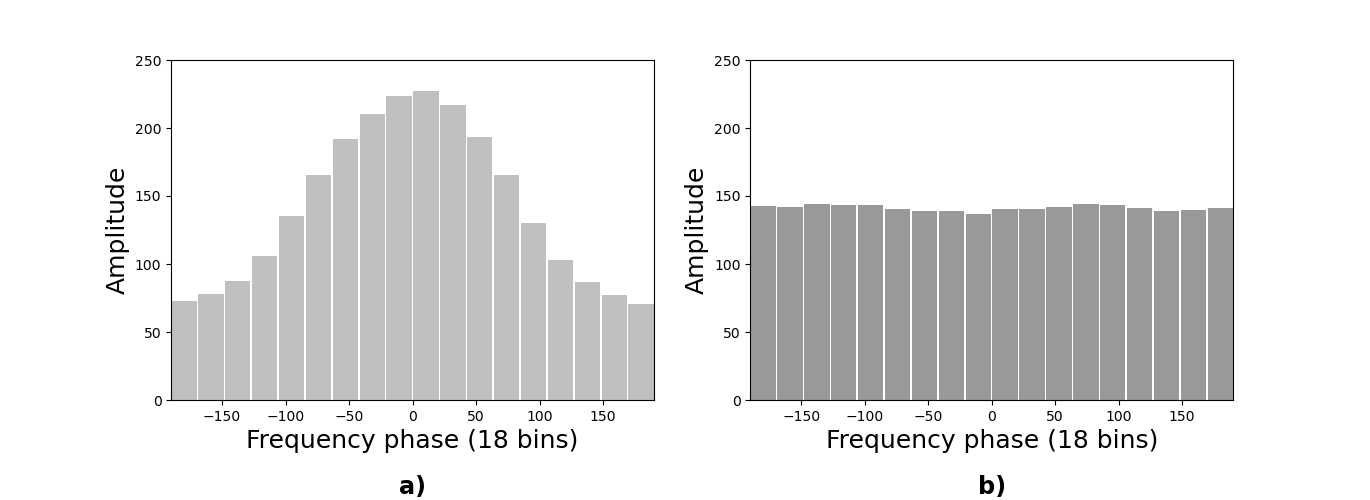}
\caption{Amplitude distribution over phase bins. a)Theta phase. b) Alpha phase. The synthetic signal contain PAC where the phase of the Theta band modulates the amplitude of the Gamma band. Thus, a) show an amplitude distribution that correspond to the presence of PAC} \label{fig2_}
\end{figure}

In order to calculate MI \cite{tort_measuring_2010}, it is necessary to extract the "phase-modulating" frequency band, $f_p$ and the "amplitude-modulated" frequency band, $f_A$, from the raw signal, $x_{raw}(t)$. After filtering $x_{raw}(t)$ at the frequency ranges of $f_p$ and $f_A$, the resulting signals are denoted as $x_{f_p}(t)$ and $x_{f_A}(t)$ respectively. Then, the time series of the phases of $x_{f_p}(t)$ and the time series of the
amplitude envelope of $x_{f_A}(t)$ are obtained, named as $\Phi_{f_p}(t)$ and $A_{f_A}(t)$. 
Afterwards, all the  phases $\Phi_{f_p}(t)$ are binned into eighteen 20 degrees intervals and the mean of $A_{f_A}(t)$ over each phase bin is computed and normalized by the following formula:
\begin{equation}
P(j)= \frac{\bar{a(j)}}{\sum^{N}_{k=1}\bar{a}(k)}
\end{equation}
where $\bar{a}$  is  the mean $A_{f_A}(t)$ value  of  the j phase  bin  and  $N=18$  is  the  total  amount  of  bins; $P$  is  referred to as the "amplitude distribution". Fig. \ref{fig2_} shows the amplitude distribution over phase bins from the synthetic signals of Fig. \ref{fig1_}. It is important to note that this amplitude distribution is not normalized as the defined mean amplitude distribution, but represents the same characteristic information.

If there is no phase-amplitude coupling, the amplitude distribution P over the phase bins is uniform. Thus, the
existence of PAC is characterized by a deviation
of the amplitude distribution P from the uniform distribution, U. Tort et al. \cite{tort_measuring_2010} defined the MI as an adaptation of the Kullback–Leibler (KL) divergence \cite{kullback_information_1951}, a metric that is widely used in statistics and in information theory to measure how one probability distribution is different from another probability distribution Q. It is defined as 

\begin{equation}
KL(P,Q)=\sum^{N}_{j=1}P(j)log\frac{P(j)}{Q(j)}
\end{equation}
To perform this adaptation, the distribution divergence measure is made to assume values between 0 and 1. The MI is therefore a constant times the KL divergence of P from the uniform
distribution. It is useful to notice that the Shannon entropy \emph{H(p)} is computed by means of

\begin{equation}
\label{eq_shannon}
H(P)=-\sum^{N}_{j=1}P(j)log{P(j)}
\end{equation}
This represents  the  inherent  amount  of  information  of  a  variable. If the Shannon  entropy  is  maximal, all the phase bins present the same amplitude (uniform distribution). The KL divergence is related to the Shannon entropy by the following formula
\begin{equation}
KL(P,U)=logN-H(P)
\end{equation}
where U is the uniform distribution. Finally, the  MI  is  calculated  by  the below formula:

\begin{equation}
MI= \frac{KL(P,U)}{logN}
\end{equation}

\begin{figure}[!htb]
\includegraphics[width=\textwidth]{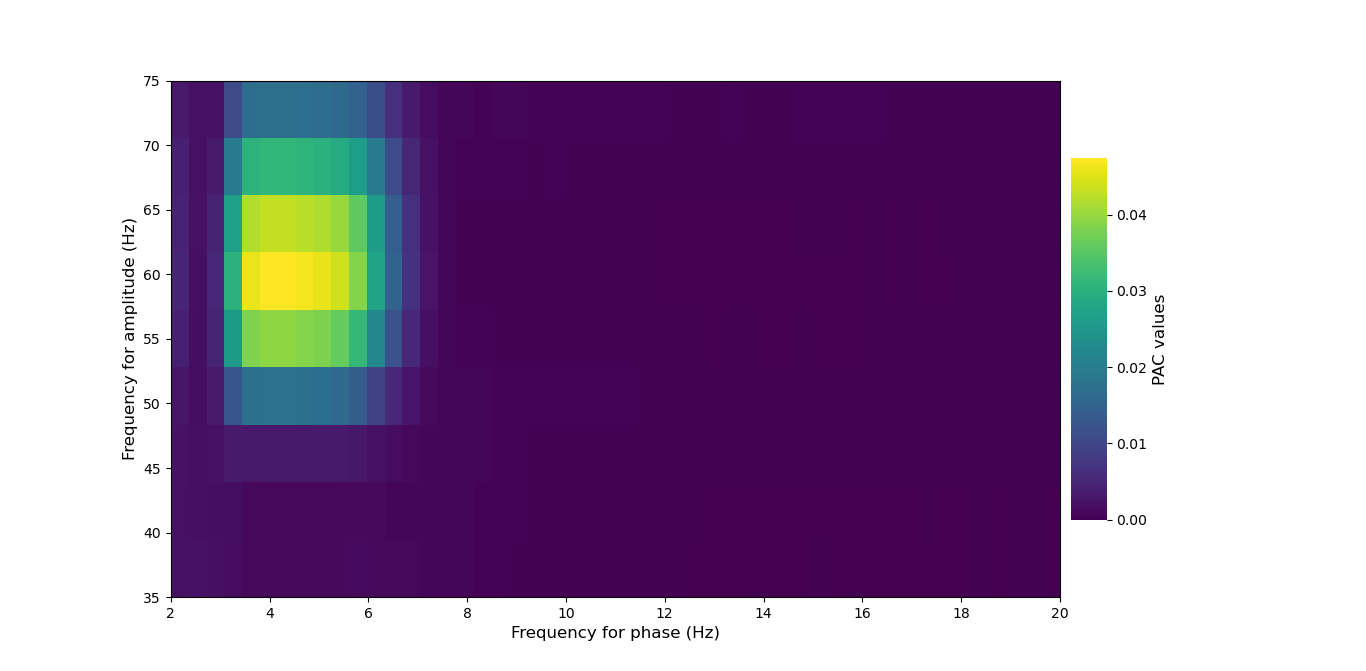}
\caption{Phase-amplitude comodulogram} \label{fig3_}
\end{figure}

The MI results were analyzed using surrogate time series, obtained from the creation of shuffled versions of $\Phi_{f_p}(t)$ and $A_{f_A}(t)$ time series. Then, this shuffled series are used to estimate a surrogate MI and this procedure is repeated several hundred times. The distribution thus obtained is considered an approximation of the null distribution \cite{hurtado_statistical_2004} and allow us to test the significance of the MI values. This null distribution allows us to infer whether the observed value actually differs from what would be expected from chance \cite{tort_measuring_2010}. To this end, 200 surrogate MI values are generated to assess the significance of the results, considering $p < 0.05$ as significant. Additionally, the MI measure is represented in a phase-amplitude comodulogram. This is a tool to represent the coupling measured among multiple frequency bands, and it is ideal to explore the PAC \cite{tort_measuring_2010}. The MI calculated for the synthetic PAC signal is shown in Fig. \ref{fig3_} where the phase of the Theta band ($f_p$) modulates the amplitude of the Gamma band ($f_A$).

To sum up, PAC is measured by MI in each temporal segment. This enables the exploration of the temporal evolution of the response to specific auditory stimuli. Thus, the PAC MI values are used to construct the nodes and edges of the coupling networks between frequency bands. To compute, PAC we used Tensorpac \cite{combrisson_tensorpac_2020}, an open-source Python toolbox dedicated to PAC analysis of neurophysiological data.

\subsection{Complex Network Analysis}

Complex networks are all around us and are present in many  scientific and technological areas, describing a wide range of real systems in nature and society. The principal tool to study these complex systems is graph theory \cite{boccaletti_complex_2006}. As a natural framework for the treatment of complex networks, it is useful to describe some common properties and how they are measured. 

\begin{figure}[h]

\begin{floatrow}

\capbtabbox{%
    \resizebox{0.5\textwidth}{!}{
    \begin{tabular}{|m{1.1cm}|m{1.1cm}|m{1.1cm}|m{1.1cm}|m{1.1cm}|m{1.1cm}|}
    \hline
     & Delta & Theta & Alpha & Beta & Gamma \\
    \hline
    Delta &  0 & 0 & 0 & 1 & 0\\
    Theta & 0 & 0 & 0 & 1 & 1 \\
    Alpha &  0 & 0 & 0 & 0 & 1\\
    Beta &  0 & 0 & 0 & 0 & 1 \\
    Gamma & 0 & 0 & 0 & 0 & 0 \\
    \hline
    \end{tabular}}
}{%
  \caption{Example of adjacency matrix for a dyslexic subject.}\label{tab1_}
}

\ffigbox{%
    \includegraphics[width=0.5\textwidth]{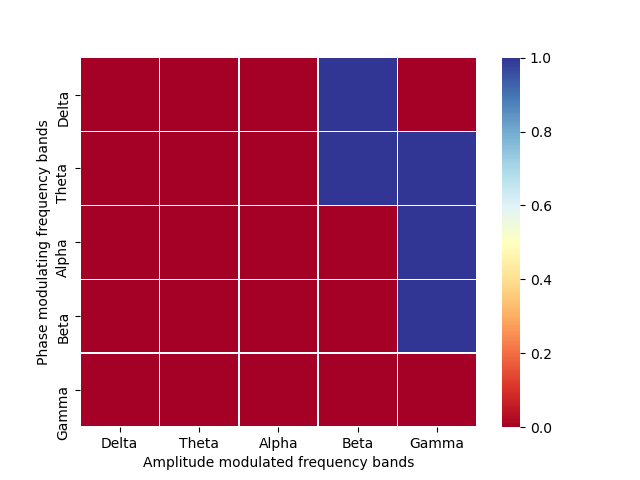}
}{%
  \caption{Adjacency matrix from Table \ref{tab1_}}\label{figMatriz}
}

\end{floatrow}
\end{figure}

A complex network is represented as a graph G(N, K), that is, two sets of N nodes and K links or edges. The graphs are undirected or directed and weighted or unweighted.  In an undirected graph, each of the links is defined by a couple of nodes $i$ and $j$, and is denoted as $(i, j)$. In a directed graph, the order of the two nodes is important: $(i, j)$ stands for a link from $i$ to $j$ and $(j, i)$ is a different link from $j$ to $i$. An unweighted graph have a binary nature,
where the edges between nodes are either present or not. In contrast, in a weighted graph each link carries a numerical value measuring the strength of the connection \cite{boccaletti_complex_2006}. If there are loops or multiple edges they are called multigraphs. A graph is completely described by its adjacency matrix, $A$. It is a $N\times N$ matrix where $a_{ij}$  with $(i, j=1,2,... N)$ is equal to 1 when the edge $l_{ij}$ exists and zero otherwise. In this work, the adjacency matrix is a $5\times 5$ asymmetric matrix from the frequency bands considered as nodes. The entry $a_{ij}$ differs from zero if there is PAC measured in an electrode between two frequency bands. This is shown in Table ~\ref{tab1_}.

The node degree, which is one of the most important descriptors for complex networks, is computed as:
\begin{equation}
k_i= \sum^{N}_{j=1}a_{ij}
\end{equation}
where $k_i$ is the degree of a node $i$ and $a_{ij}$ is an entry for the adjacency matrix.  This way, considering the number of connections that link it to the rest of the network. The degree distribution is drawn from the degree of all nodes. Another measure directly related to the node degree is the assortativity. It is the correlation between the degrees of connected nodes \cite{bullmore_complex_2009}. Furthermore, the relation between the actual number of edges and the number of possible edges of a graph provide the density as
\begin{equation}
D = \frac{K}{N(N-1)}
\end{equation}
for directed graphs.

The clustering coefficient $C$ is defined as follows \cite{watts_collective_1998}. Suppose that a node $i$ has $k_i$ neighbors; then at most $k_i(k_i-1)/2$ edges exist between them (this occurs when every neighbour of $i$ is connected to every other neighbor of $i$). Let $C_i$ denote the fraction of these allowable edges that actually exist. Define $C$ as the average of $C_i$ over all $i$. 
\begin{equation}
C= \frac{1}{N}\sum^{N}_{i=1}C_i
\end{equation}
where N is the total number of vertex and $C_i$ is the clustering coefficient of vertex i

Path length, L, is defined in \cite{watts_collective_1998} as the number of edges in the shortest path between two vertices, averaged over all pairs of vertices. Also, we have that

\begin{equation}
L= \frac{1}{N}\sum^{N}_{i=1}L_i
\end{equation}
where $L_i$ is the average distance between vertex i and
all other vertex.

The centrality, in general terms, quantifies the relative relevance of a given node in a network. The node betweenness measures how many of the shortest paths between all other node pairs in the network pass through a node \cite{bullmore_complex_2009}. The degree, closeness and betweenness are the standard measures of node centrality. 

The study of real networks has shown the existence of bridging links that connect different areas of most of the networks. This means that there is a short path between any two nodes that speed up the communication among nodes \cite{boccaletti_complex_2006}.
This particular organization is known as the small-world property or small-worldness. Networks with such topology are named small world networks and are typified by having a small value of $L$, like random graphs, and a high clustering coefficient $C$, like regular lattices than would be expected by chance \cite{watts_collective_1998,cohen_analyzing_2014}. The small-worldness of a network is measured by comparing it to an Erdös-Rényi (E-R) \cite{erdos_evolution_2011} random graph with the same nodes and edges. For this purpose, a quantitative metric of small-worldness, $S$, is defined as
\begin{equation}
S= \frac{\gamma}{\delta}=\frac{\frac{C}{C_{rand}}}{\frac{L}{L_{rand}}}
\end{equation}
where $L$ is the mean shortest path length of a network and $C$ its clustering coefficient. $L_{rand}$ and $C_{rand}$ are the respective quantities for the corresponding E–R random graph. From \cite{humphries_network_2008} a network is said to be a small-world network if $L\geq L_{rand}$ and $C\gg C_{rand}$. Thus, a network is said to be a small-world network if $S>1$.

\subsubsection{Network Construction} 
\begin{figure}[!htb]
\includegraphics[width=\textwidth]{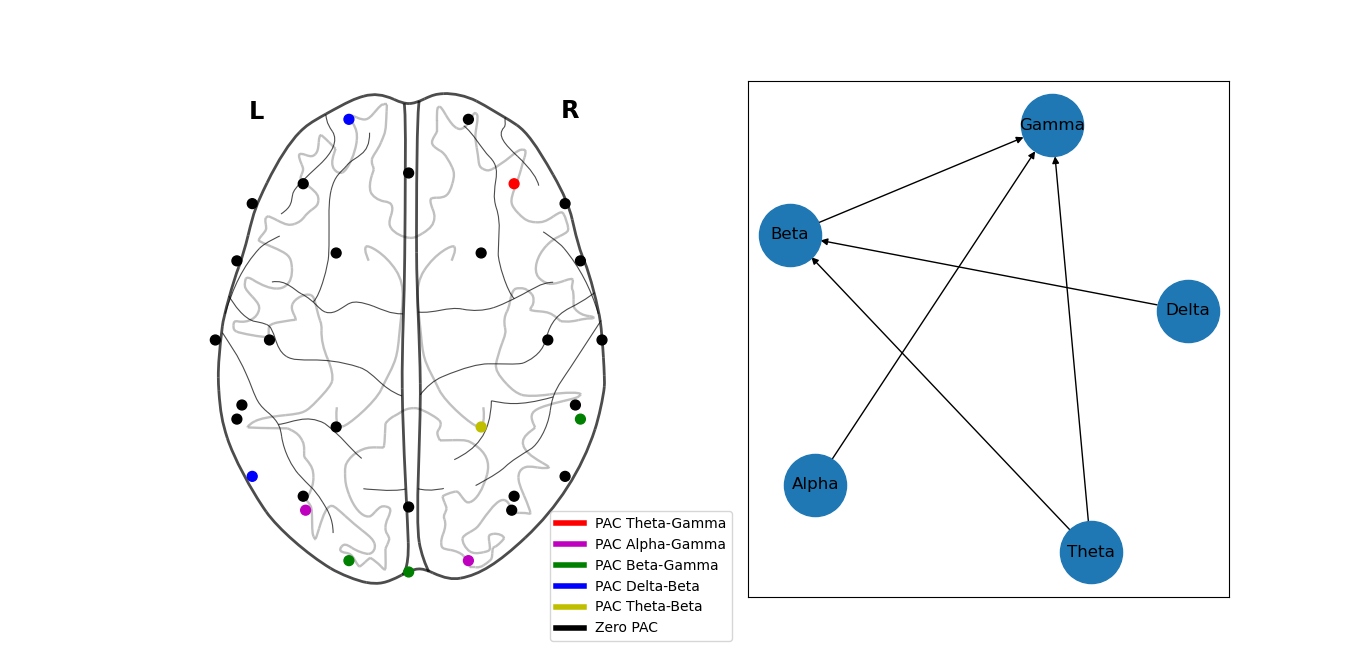}
\caption{Frequency bands source for PAC and graph from adjacency matrix in Table \ref{tab1_}  } \label{fig4_}
\end{figure}

There are numerous works \cite{strogatz_exploring_2001,reka_statistical_2002,newman_structure_2003} about the study of characteristics of complex networks. It is defined as a network with certain topological features such as high clustering, small-worldness, the presence of high-degree nodes or hubs, assortativity, modularity or hierarchy. This does not occur or are not typical of simple networks such as random graphs or regular lattices \cite{bullmore_complex_2009}. The complex network used in this work are based in the MI explained in the above section. In this way, we employed directed and unweighted graphs with the nodes being the typical EEG frequency bands in which the PAC is measured. Thus, the edges of the graphs correspond to the presence of PAC in an electrode between a pair of frequency bands with a direction from a band providing the phase to another band that its amplitude is modulated. These edges are binarized, therefore, unweighted. From the adjacency matrix in Table \ref{tab1_} it is obtained the graph shown in Fig. \ref{fig4_}. Each entry of the adjacency matrix equal to 1 is due to a significant PAC MI value in an electrode for a determined combination of frequency bands. This is represented on the left part of Fig. \ref{fig4_}. 
Then, graphs are obtained for each subject and each temporal segment, enabling the study of the network characteristics through time. This has been done with the use of the Python package NetworkX that is meant for the creation, manipulation, and study of the structure, dynamics, and functions of complex networks \cite{hagberg_exploring_2008}.

\section{Experimental Results}
\label{sec:Results}
In this section, we present the results for the PAC analysis, complex network analysis and classification. First, we start with the results of the PAC analysis over the temporal segments that allow the exploratory analysis. To this end, the MI measured intra-electrode is represented for each frequency band combination considered. With this, extracting information about the temporal evolution of the PAC for dyslexic and control subjects. Then, we focus on the complex network analysis, obtaining metrics to study the networks and the evolution of its characteristics through time. Finally, these metrics are used as features to train a SVM classifier to perform a classification test.

\subsection{PAC Results} The PAC is measured and presented as a function of temporal segments. A set of frequency band pairs are defined in tensorpac to study and identify a temporal behavior by measuring the MI. These combinations of frequency bands are: Delta-Gamma, Theta-Gamma, Alpha-Gamma, Beta-Gamma, Delta-Beta, Theta-Beta and Alpha-Beta. We have measured the PAC for each subject and each temporal segment, obtaining results for all the combinations of frequency bands.

\begin{figure}[!htb]
\includegraphics[width=\textwidth]{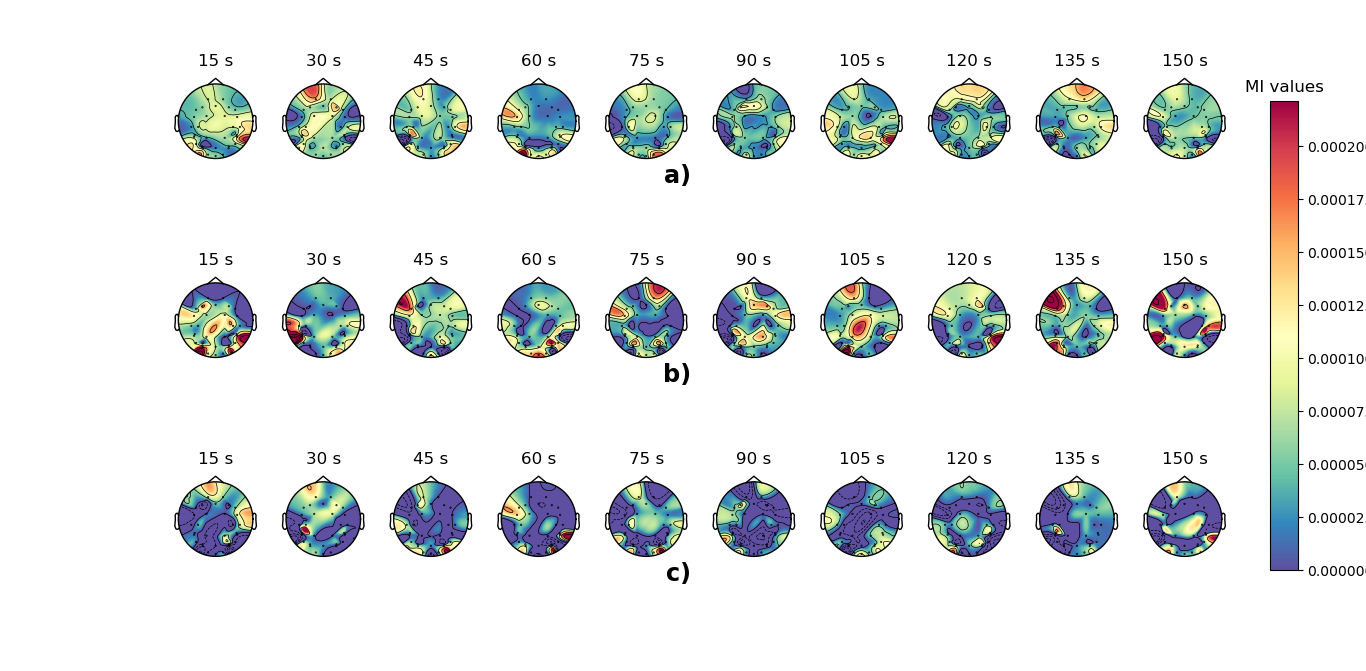}
\caption{Average MI topoplots for Theta and Beta frequency bands and 4.8 Hz stimulus. a) Controls b) Dyslexics c)Controls and dyslexics difference } \label{fig1}
\end{figure}

\begin{figure}[!htb]
\includegraphics[width=\textwidth]{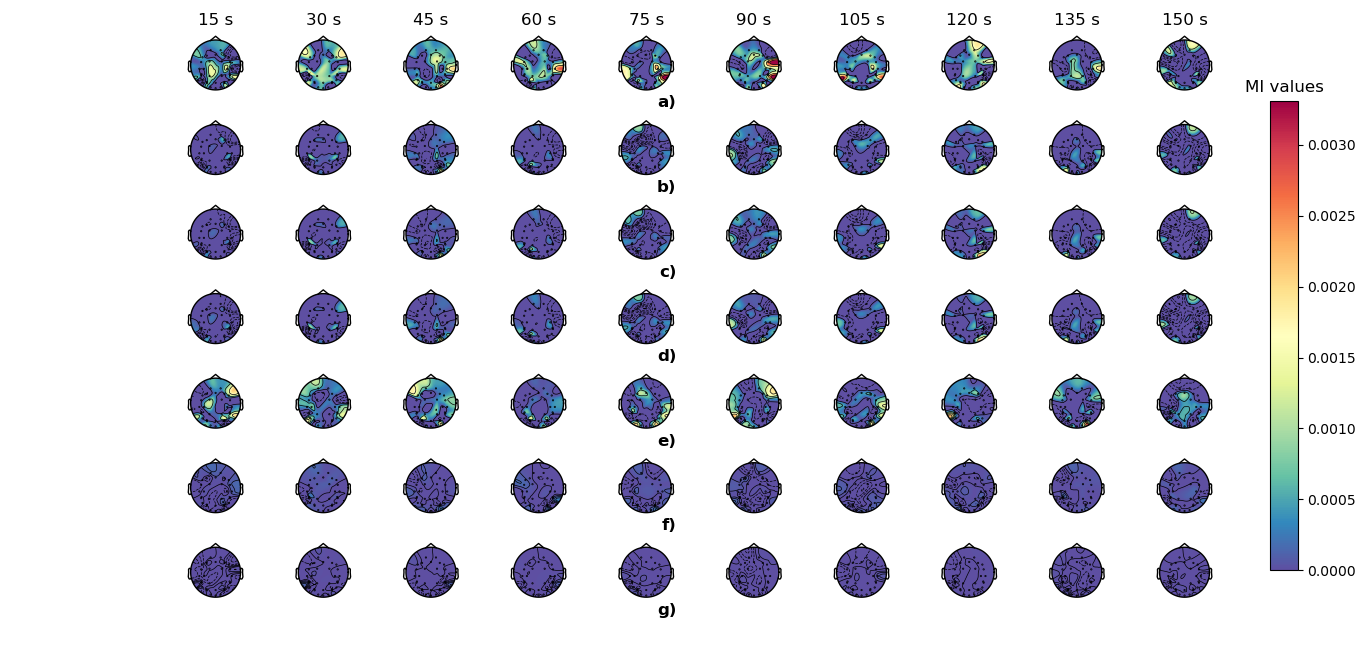}
\caption{Difference average MI topoplots for 4.8 Hz. a) Delta-Gamma b) Theta-Gamma c) Alpha-Gamma d) Beta-Gamma e) Delta-Beta f) Theta-Beta g) Alpha-Beta} \label{fig2}
\end{figure}

The MI measured in each electrode is represented with a set of ten topoplots showing the temporal evolution of the average PAC for dyslexic and control subjects in every frequency combination. Fig. \ref{fig1} shows the average MI measured for each segment and every subject between the Theta and Beta frequency bands. In Fig. \ref{fig1} a) and b) the MI for controls and dyslexics are presented, respectively. The differences in the temporal evolution is easier to see and interpret in Fig. \ref{fig1} c) in which the topoplots show the differences between the average MI value for the dyslexic group and the control group. The topoplots in Fig. \ref{fig2} show the difference between controls and dyslexics for the MI measured in the rest of frequency bands combinations.

\subsection{Complex Network Analysis Results}

\begin{figure}[!htb]
\includegraphics[width=\textwidth]{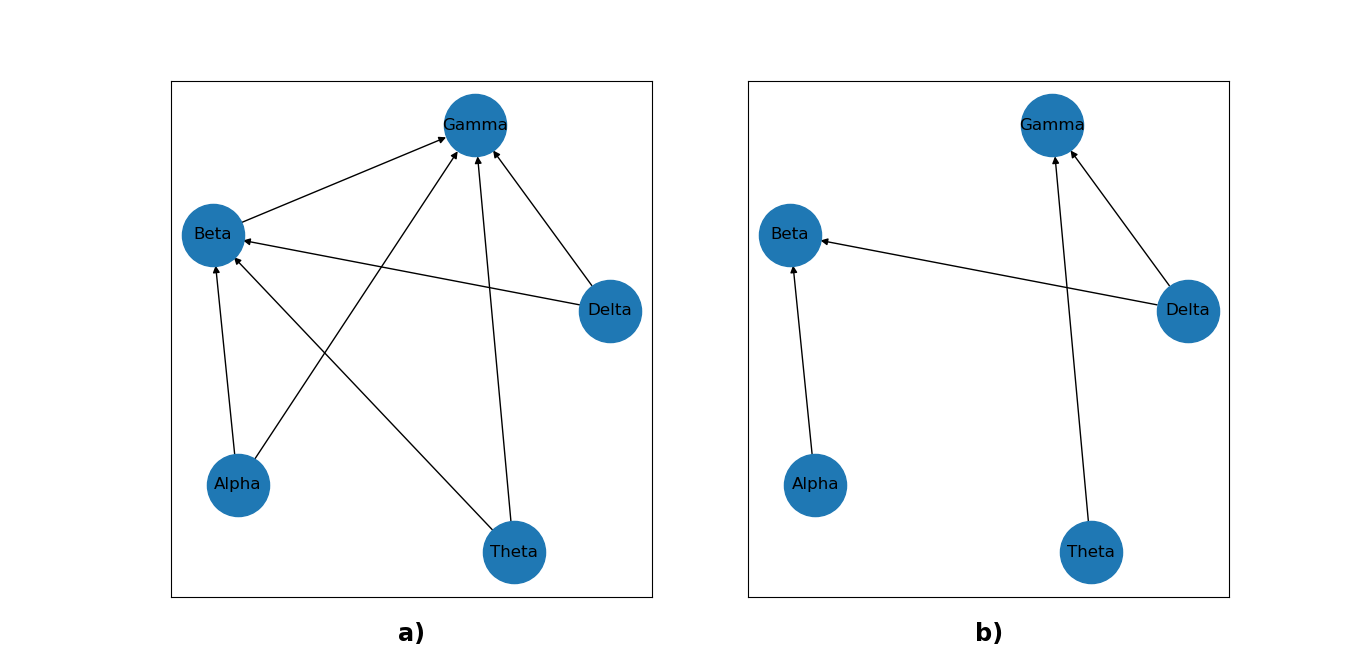}
\caption{Graphs showing the presence of MI intra-electrode PAC in a pair of EEG bands. a)Control subjects. b) Dyslexic subjects} \label{fig3}
\end{figure}

In this section, the selected characteristics of the complex networks that have been measured are presented. From the MI results of the PAC analysis, a set of ten graphs (one for each temporal segment) have been created for each subject. To clarify, each graph contains the information of the measured MI of all electrodes for a temporal segment of one subject. Thus, as we can see in Fig. \ref{fig3} the edges in the graph represents the existence of measured PAC in an electrode. These edges appear from a node indicating the "phase modulating" frequency band, to the pointed node representing the "amplitude modulated" frequency band. 
There is a great variability between subjects and segments in the number of nodes and edges, with a trend of having less edges in certain temporal segments in dyslexic subjects.

\begin{figure}[!htb]
\includegraphics[width=\textwidth]{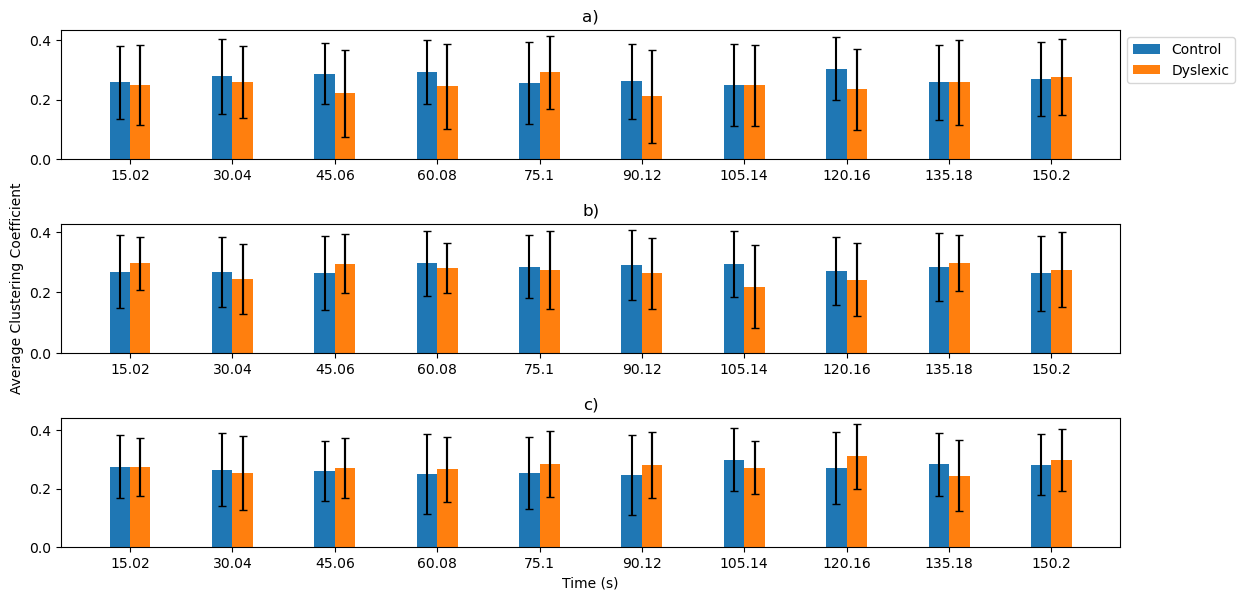}
\caption{Average clustering coefficient. a)4.8 Hz b)16 Hz c)40 Hz} \label{fig5}
\end{figure}

\begin{figure}[!htb]
\includegraphics[width=\textwidth]{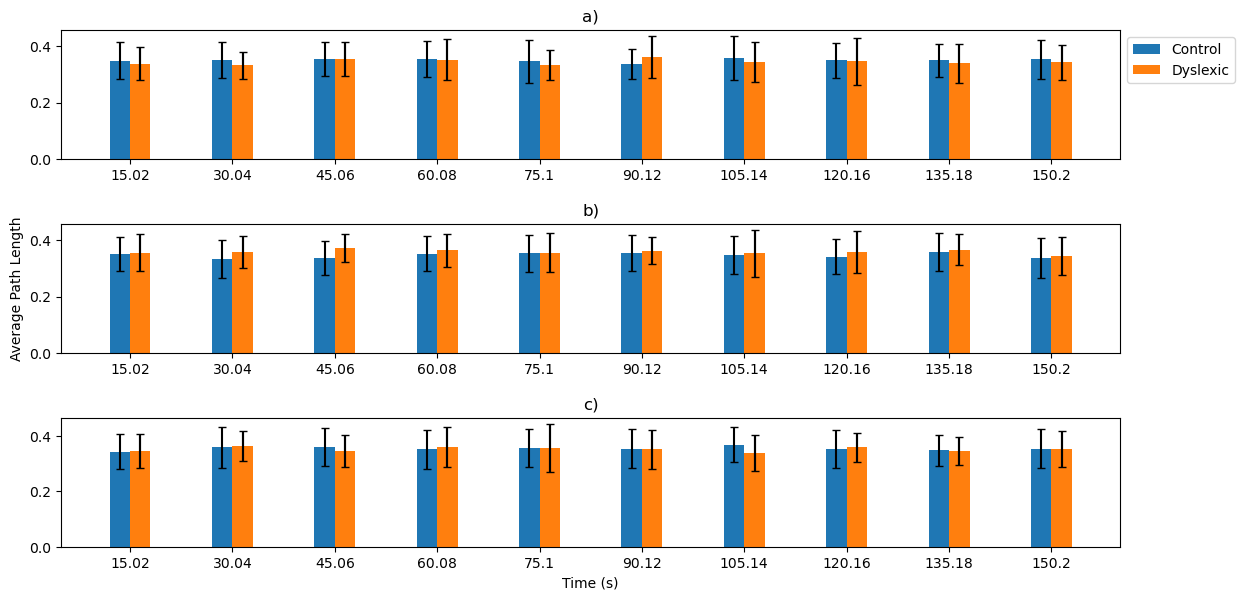}
\caption{Average path length. a)4.8 Hz b)16 Hz c)40 Hz} \label{fig6}
\end{figure}

\begin{figure}[!htb]
\includegraphics[width=\textwidth]{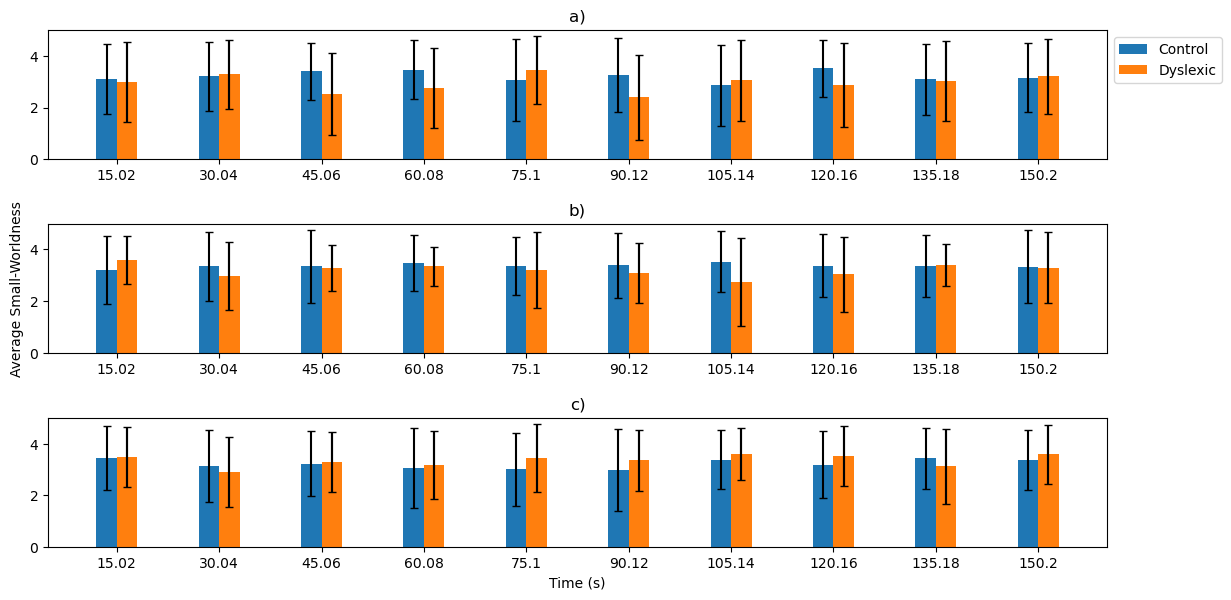}
\caption{Average small-worldness. a)4.8 Hz b)16 Hz c)40 Hz} \label{fig8}
\end{figure}

Once the graphs are calculated, we measured the selected metrics. First, the segregation of the networks is characterized with the clustering coefficient. In Fig. \ref{fig5} the average clustering coefficient of each graph for 4.8 Hz, 16 Hz and 40 Hz is shown.  
Second, the characteristic path length is measured obtaining the average path length of each graph and represented it in Fig. \ref{fig6}. We see that the average path length is similar in most cases for both groups, but in certain temporal instants the clustering coefficient is greater for the 4.8 Hz graphs in the control group.
These metrics make possible to calculate the small-worldness of the complex networks for each subject and segment. This is shown in Fig. \ref{fig8} where, as expected from the measures of C and L, the graphs in 4.8 Hz have a consistent small-world topology for control subjects, while the dyslexic graphs lost this property in some temporal segment. For 16 and 40 Hz these differences are not so apparent.

\subsection{Classification}

Here, the metrics obtained from the analysis of the characteristics of complex networks are used as features to train a SVM \cite{cortes_supportvector_1995} classifier. In particular, the classification results from the small-worldness property are presented and discussed as it achieves a better performance. Using the small-worldness data, a $NxM$ matrix  has been created with $N$, the number of subjects, and $M$, the number of segments in which the small-worldness is measured. To evaluate this, the metrics used in the classification are the accuracy, sensitivity, specificity and Area Under ROC Curve (AUC). Then, cross validation is applied to the data using a K-fold stratified scheme with 5 folds. 

\begin{figure}[!htb]
\includegraphics[width=\textwidth]{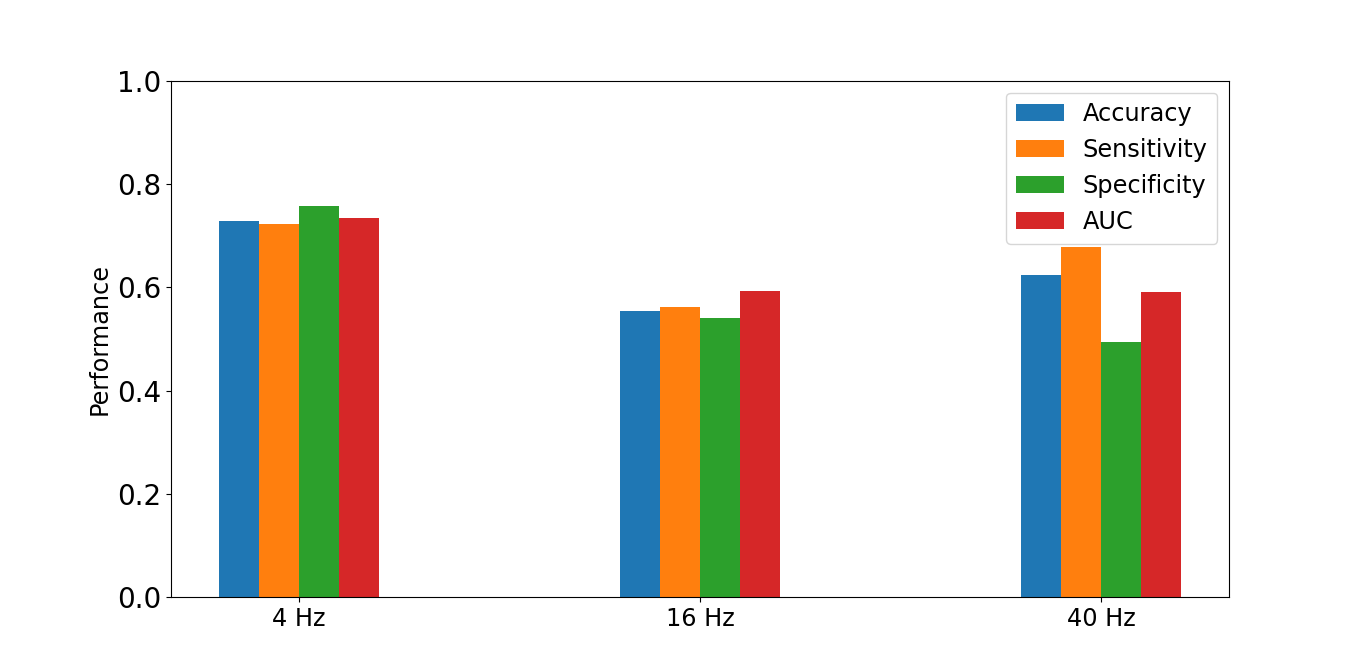}
\caption{Classification results for 4.8 Hz, 16 Hz and 40 Hz.} \label{fig9}
\end{figure}

As it is exposed in the above section, the graphs obtained for the 4.8 Hz stimulus PAC measured implied a greater difference in the small-worldness between groups. This is traduced to a better classification performance for 4.8 Hz stimulus using the small-worldness data that is shown in Fig. \ref{fig9}. In this figure the accuracy, sensitivity, specificity and AUC are represented and in the Table \ref{tab1} metrics for other features are shown.

\begin{table}
\centering
\caption{Classification comparative using graph metrics as SVM features.}\label{tab1}
\begin{tabular}{|m{2cm}|m{2cm}|m{2cm}|m{2cm}|m{2cm}|m{2cm}|}
\hline
Stimulus & Features & Accuracy & Sensitivity & Specificity & AUC\\
\hline
4.8 Hz & Small-Worldness &  \bf{0.729} & \bf{0.723} & \bf{0.747} & \bf{0.733}\\
 & Average  Degree & 0.519 & 0.467 & 0.647 & 0.575 \\
 & Assortativity &  0.489 & 0.4 & 0.71 & 0.626 \\
 & Density &  0.372 & 0.23 & 0.727 & 0.518 \\
 & Betweenness Centrality &  0.522 & 0.458 & 0.6 & 0.595 \\
\hline
16 Hz & Small-Worldness &  0.554  & 0.5612 & 0.54 & 0.593 \\
 & Degree & 0.438 & 0.445 & 0.433 & 0.426 \\
 & Assortativity &  0.511 & 0.354 & 0.893 & 0.629 \\
 & Density &  0.573 & 0.632 & 0.433 & 0.524 \\
 & Betweenness Centrality &  0.372 & 0.2 & 0.8 & 0.659 \\
 \hline
 40 Hz & Small-Worldness & 0.624 & 0.678 & 0.493 & 0.591 \\
 & Degree & 0.5 & 0.512 & 0.473 & 0.53 \\
 & Assortativity &  0.371 & 0.2 & 0.8 & 0.626 \\
 & Density &  0.564 & 0.559 & 0.573 & 0.596 \\
 & Betweenness Centrality &  0.372 & 0.2 & 0.8 & 0.555 \\
 \hline
\end{tabular}
\end{table}

\begin{figure}[!htb]
\includegraphics[width=\textwidth]{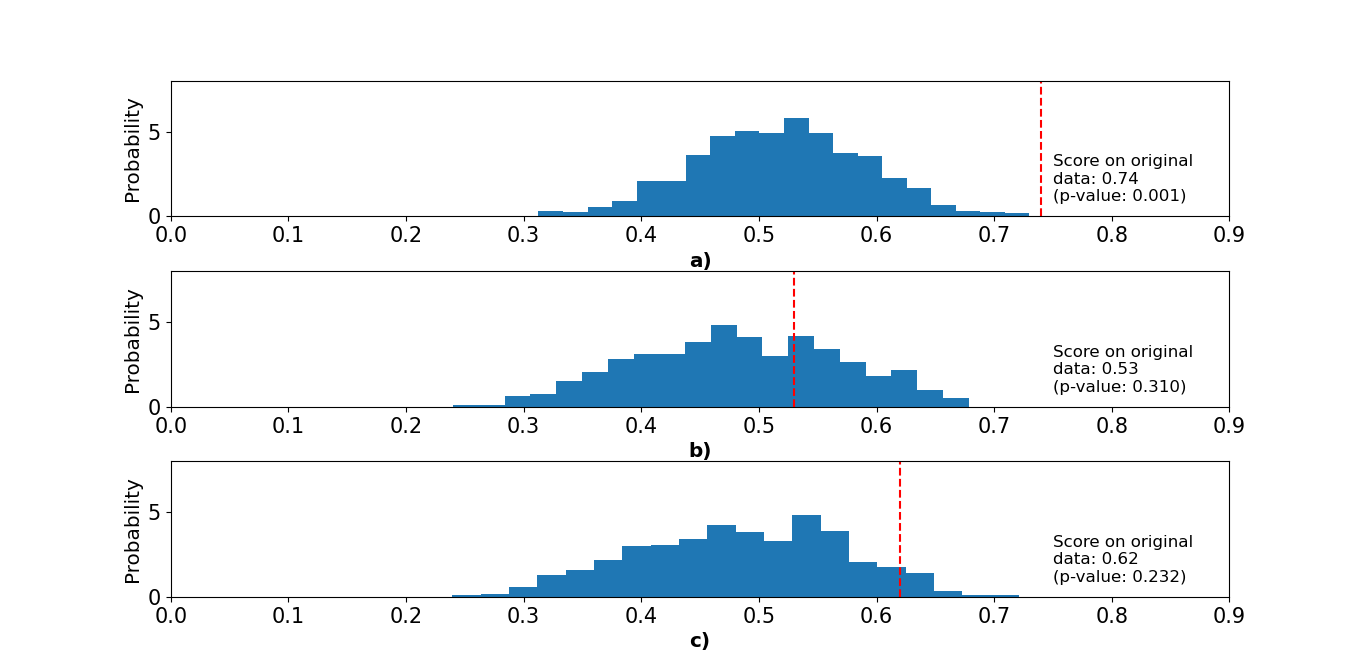}
\caption{Results of the permutation tests using the small-worldness property as a feature. a)4.8 Hz b)16 Hz c)40 Hz} \label{fig10}
\end{figure}

At this point, it is important to estimate the significance of the classification accuracy obtained and evaluate whether the classifier has found a real connection between the data and the class labels \cite{ojala_permutation_2009}. Permutation tests are used for this purpose, providing useful statistics about the underlying reasons for the obtained classification result. Thus, under the null hypothesis that the features and the labels are independent the null distribution is estimated by 1000 permutations of the data set labels in a 5-fold cross validation scheme. The p-values obtained represent the fraction of random data sets where the classifier behaved as well as or better than in the original data. Fig. \ref{fig10} illustrates the accuracy obtained by the random data and the original score along with the corresponding p-value for each case. The p-value is only lower than 0.05 for the 4.8 Hz data. Consequently,  a significant accuracy score is obtained and is greater than for the random data.

\section{Discussion}
\label{sec:Discussion}
The application of graph theory to complex networks is a promising area to study the brain and its pathologies. In the present work, we have performed an exploratory analysis using complex networks that are extracted from PAC. This PAC has been measured between the typical EEG frequency bands: Delta, Theta, Alpha, Beta and Gamma. All the effort is oriented to advance towards an early Developmental Dyslexia diagnosis and a better understanding of its neural basis. There are other works that have studied this neurological disorder to develop machine learning methods for detecting Dyslexia as in Table \ref{tab2} using EEG, fMRI and MRI. For this, it is fundamental to count with a database that offer adequate EEG data from a sufficient number of control and dyslexics subjects. In this case, the database is provided by the Leeduca Study Group at the University of Málaga. It is obtained from the experiment where a set of three auditory stimulus 4.8, 16 and 40 Hz is presented to controls and dyslexics. According to Goswami \cite{goswami_temporal_2011} this auditory stimulus are related to the oscillatory  phase  entrainment to speech in brain that are thought to be important for speech encoding. Furthermore, it is reinforced in \cite{goswami_neural_2018} the existence of phonological impairments in all linguistic levels in dyslexics children such as stressed 
syllables, syllables, onset-rimes, and phonemes. The last decade has faced a change of the approach to language processing with new tools and ideas. This was stated in Poeppel \cite{poeppel_neuroanatomic_2014}, where it is pointed out that the system for speech and language processing is considerably more complex and distributed. 

Neural oscillations have become a promising mechanism for studying the processes of learning and language. The MI as defined by Tort et al. \cite{tort_measuring_2010} has been used to obtain the PAC intra-electrode for each subject and for every temporal segment specified. The PAC results are shown in topoplots that represents the electrode activation in terms of the MI values. The results for the Theta-Beta PAC in the 4.8 Hz stimulus are presented in Fig.~\ref{fig1} as an example of the outcome of the PAC analysis. This figure shows the evolution of the average MI for both control and dyslexic groups (Fig.~\ref{fig1} a) and Fig.~\ref{fig1} b), respectively) through each temporal segment and in every electrode. To visualize the distinction between groups, it is represented the difference of the mean MI as in Fig.~\ref{fig1} c). There is a high variability in the topoplots of each subject and that makes it difficult to perform a visual analysis of the differences. To overcome this, the difference average MI for each frequency band combinations is represented in Fig.~\ref{fig2} for the 4.8 Hz stimulus, showing larger differences for the temporal segments of the PAC from Delta-Gamma and Delta-Beta. This implies that for these bands the response to the auditory stimulus differ over time for controls and dyslexics.

A set of complex networks have been created from the PAC measured. 
These networks represent the interaction between frequency bands through time by means of intra-electrode PAC. An example of this is shown in Fig.~\ref{fig3}. Each graph shows the results of PAC between every pair of frequency bands for each subject and each temporal segment, highlighting the presence of PAC in one or several electrodes. Among all the subjects the graph obtained are diverse, although,  there is a trend to have less edges for dyslexics graphs. This suggests a lower number of electrodes with significant PAC measured.

Once the graphs are obtained, the graph metrics are calculated to study the characteristics and its evolution through time. From all the results it is highlight the property of small-worldness (Fig. \ref{fig8}), denoting a small-world topology in agreement with other brain complex networks as stated in \cite{sporns_small_2006}. Works as \cite{sporns_small_2006,bullmore_complex_2009} provide strong evidence that brain networks exhibit small-world attributes and this is preserved across multiple frequency bands. Regarding the small-worldness property, the average clustering coefficient and the average path length for the three stimulus are shown in Fig.~\ref{fig5} and Fig. \ref{fig6}, respectively. From these results, it is noted that dyslexics subjects have a lower clustering coefficient in certain moments of the temporal response for the 4.8 Hz auditory stimulus that cause a reduction of the small-world topology and, thus, a loss of efficiency in the communication of the complex network.

Finally, in Table \ref{tab1} there are presented the classification results for all the metrics studied in this work. From this, we assess the classification performance of every metric and concluded that the small-worldness property has a larger discrimination capacity. The best classification results are for the 4.8 Hz stimulus with the small-worldness property achieving an accuracy of 0.729, 0.723 sensitivity,  0.747 specificity  and an AUC of 0.733 as shown in Fig. \ref{fig9}. For all the stimulus the small-world property provide with better classification results than other metrics. Then, a permutation test is performed to estimate the significance of the classification accuracy obtained. Fig. \ref{fig10} indicates that the p-value is only lower than 0.05 for the 4.8 Hz small-worldness data, thus, rejecting the null hypothesis and finding a real connection between the data and the class labels. 

\begin{table}

\begin{center}
\resizebox{\textwidth}{!}{
\caption{Classification results obtained in different works using structural imaging (MRI), functional imaging (fMRI) and electroencephalography (EEG). \\ (*) Data not available in the source.}\label{tab2}
\begin{tabular}{|m{2cm} m{2cm} m{2cm} m{2cm} m{2cm} m{2.5cm}|}
\hline

\multirow{3}{2cm}{\centering Author} & \multirow{3}{2cm}{\centering Acquisition Technique}  & \centering Number &  \multirow{3}{2cm}{\centering Stimulus} & \centering Machine   & \multirow{3}{2cm}{\centering Classifier Performance} \\
 &  & \centering of & &\centering Learning & \\
\centering & &  \centering electrodes & &\centering Method & \\

\hline

\multirow{5}{2cm}{\centering Cui et al.\cite{cui_disrupted_2016}} &
\multirow{5}{2cm}{\centering MRI} & \multirow{5}{2cm}{\centering -} & \multirow{5}{2cm}{\centering Reading} & \multirow{5}{2cm}{\centering SVM} & Acc= 0.836$\pm$*\\ & & & & & Sens= 0.75$\pm$*\\ & & & & & Spec= 0.909$\pm$*\\ & & & & &  AUC=  0.86$\pm$*\\& & & & & \\


\multirow{5}{2cm}{\centering Płoński et al.\cite{plonski_multiparameter_2017}} & \multirow{5}{2cm}{\centering MRI} & \multirow{5}{2cm}{\centering -}  & \multirow{5}{2cm}{\centering Reading} & \centering SVM & Acc= 0.65$\pm$*\\ & & & &\centering LR & Sens= *\\ & & & &\centering RF & Spec= *\\ & & & & &  AUC=  0.66$\pm$*\\& & & & & \\


\multirow{3}{2cm}{\centering Frid and Manevitz\cite{frid_features_2019}} & \multirow{3}{2cm}{\centering EEG} &  \multirow{4}{2cm}{\centering 64} &  \multirow{3}{2cm}{\centering Reading} & \multirow{4}{2cm}{\centering SVM} & \multirow{3}{2cm}{\centering Confusion matrix}\\& & & & &\\& & & & &\\


\multirow{4}{2cm}{\centering Perera et al.\cite{perera_eeg_2018}} & \multirow{4}{2cm}{\centering EEG} &  \multirow{4}{2cm}{\centering 32} & \multirow{4}{2cm}{\centering Typing and writing} & \multirow{4}{2cm}{\centering SVM} & Acc= 0.718$\pm$*\\ & & & & & Sens= 0.764$\pm$*\\ & & & & & Spec= 0.667$\pm$*\\ & & & & & AUC=  0.86$\pm$*\\& & & & & \\


\multirow{4}{2cm}{\centering Rezvani et al.\cite{rezvani_machine_2019}} & \multirow{4}{2cm}{\centering EEG} &  \multirow{4}{2cm}{\centering 64} & \multirow{4}{2cm}{\centering Typing and writing} & \multirow{4}{2cm}{\centering SVM} & Acc= 0.953$\pm$*\\ & & & & & Sens= 0.964$\pm$*\\ & & & & & Spec= 0.933$\pm$*\\& & & & & AUC= *\\& & & & & \\


\multirow{4}{2cm}{\centering Frid and Breznitz \cite{frid_svm_2012}} & \multirow{4}{2cm}{\centering EEG} &  \multirow{4}{2cm}{\centering 64} &  \multirow{4}{2cm}{\centering Auditory} & \multirow{4}{2cm}{\centering SVM} & Acc=0.85$\pm$0.1\\ & & & & & Sens= 0.749$\pm$0.2\\ & & & & & Spec=0.822$\pm$0.12\\& & & & & AUC= *\\& & & & & \\


\multirow{4}{2cm}{\centering Chimeno et al.\cite{garciachimeno_automatic_2014}} & \multirow{4}{2cm}{\centering fMRI} &  \multirow{4}{2cm}{\centering 32} & \multirow{4}{2cm}{\centering Tests} & \multirow{4}{2cm}{\centering ANN} & Acc=0.949$\pm$*\\ & & & & & Sens= 0.947$\pm$*\\ & & & & & Spec=0.95$\pm$*\\& & & & & AUC= *\\& & & & & \\


\multirow{4}{2cm}{\centering Zahia et al.\cite{zahia_dyslexia_2020}} & \multirow{4}{2cm}{\centering fMRI} &  \multirow{4}{2cm}{\centering 32} &  \multirow{4}{2cm}{\centering Reading} & \multirow{4}{2cm}{\centering 3D CNN} & Acc=0.727$\pm$*\\ & & & & & Sens= 0.75$\pm$*\\ & & & & & Spec=0.714$\pm$*\\& & & & & AUC= *\\& & & & & \\


\multirow{4}{2cm}{\centering Ortiz et al.\cite{ortiz_anomaly_2019}} & \multirow{4}{2cm}{\centering EEG} &  \multirow{4}{2cm}{\centering 32} &  \multirow{4}{2cm}{\centering Auditory} & \multirow{4}{2cm}{\centering SVM} & Acc=0.78$\pm$*\\ & & & & & Sens= 0.66$\pm$*\\ & & & & & Spec=0.81$\pm$*\\ & & & & & AUC= 0.83$\pm$*\\& & & & & \\


\multirow{4}{2cm}{\centering Martinez-Murcia et al.\cite{martinez-murcia_eeg_2020}} & \multirow{4}{2cm}{\centering EEG} &  \multirow{4}{2cm}{\centering 32} &  \multirow{4}{2cm}{\centering Auditory} & \multirow{4}{2cm}{\centering Autoencoder} & Acc=0.74$\pm$0.114\\ & & & & & Sens= 0.596$\pm$0.254\\ & & & & & Spec=0.79$\pm$0.221\\ & & & & & AUC= 0.762$\pm$*\\& & & & & \\


\multirow{5}{2cm}{\centering Martinez-Murcia et al.\cite{martinez-murcia_periodogram_2019}} & \multirow{4}{2cm}{\centering EEG} &  \multirow{4}{2cm}{\centering 32} &  \multirow{4}{2cm}{\centering Auditory} & \multirow{4}{2cm}{\centering SVM} & Acc=0.728$\pm$*\\ & & & & & Sens= 0.667$\pm$*\\ & & & & & Spec=0.789$\pm$*\\ & & & & & AUC=0.748$\pm$*\\
\hline
\end{tabular}}
\end{center}
\end{table}

\clearpage

\section{Conclusions}
\label{sec:Conclusions}
The small-world attributes are present in the majority of functional brain networks. This has been a starting point to this work, with the concept that network organization in neurological disease reflects a deviation from the optimal pattern. Even though the complex networks studied in this work are not directly functional brain networks, they are obtained from PAC brain coupling for each electrode under an experimental setup using non-interactive and non-speech auditory stimulus. This corresponds to the sampling processes developed in the brain during language processing and consist of bandwidth-limited white noise modulated in amplitude with 4, 16 and 40-Hz signals. Moreover, the networks provide information related to the frequency bands and temporal evolution of the PAC in the electrodes.

Graph theoretical analysis reveals abnormal patterns of organization that correspond to neurological disorders. The complex networks studied present small-world characteristics as having a high clustering coefficient and a small value of path length. However, the graphs have different pattern topology over time for control and dyslexic subjects. This underline that dyslexic graphs lose the small-world topology over the duration of the stimulus, presenting a lower clustering coefficient. After graph analysis, this metrics were used as SVM features in order to classify the subjects. As a result, the small-world property provides discriminant information to achieve AUC values up to 0.733 and obtaining the best accuracy, sensitivity and specificity for the 4.8 Hz stimulus.  

It is worth noting that the main aim of this work is to provide a
methodology for the exploratory analysis of the brain processes involved in low-level auditory processing, analysed from a coupling point of view and modelled by complex network analysis. However, the results obtained show that complex network based features are able to differentiate between controls and dyslexic subjects with similar sensitivity and specificity values that other discriminative methods (Table \ref{tab2}). In future works, we plan to extend the current study by combining it with functional brain networks. This should help to achieve better results relevant to the diagnosis and early detection of DD. This is performed by measuring PAC inter-electrode, obtaining complex networks, where it is more direct to extract information about the underlying neurocognitive profiles in DD. Another possible way, is to improve the extraction of the phase and amplitude time series with the use of adaptive decomposition methods as Empirical Mode Decomposition (EMD).

\section*{Acknowledments}
This work was supported by projects PGC2018-098813-B-C32 (Spanish “Ministerio de Ciencia, Innovación y Universidades”), and by European Regional Development Funds (ERDF) and BioSiP (TIC-251) research group. We gratefully acknowledge the support of NVIDIA Corporation with the donation of one of the GPUs used for this research. Work by F.J.M.M. was supported by the MICINN ``Juan de la Cierva - Incorporaci\'on'' Fellowship. We also thank the \textit{Leeduca} research group and Junta de Andalucía for the data supplied and the support. Funding for open access charge: Universidad de Málaga / CBUA.

 \bibliographystyle{elsarticle-num} 
 \bibliography{paper}





\end{document}